\documentclass[%
 reprint,
 amsmath,amssymb,
 aps,
pre,showkeys 
]{revtex4-1}
\usepackage{CJK}
\usepackage{graphicx}
\usepackage{dcolumn}
\usepackage{bm}

\begin{document}

\title{ Electrical percolation in metal wire network based strain sensors }

\author{Ankush Kumar}
\affiliation{Chemistry and Physics of Materials Unit, Jawaharlal Nehru Centre for Advanced Scientific Research,560064, Bangalore, India}

\email{ankush.science@gmail.com}
\altaffiliation {Present Address: Department of Mathematics, University of Pittsburgh.}

\begin{abstract}
Metal wire networks rely on percolation paths for electrical conduction, and by suitably introducing break-make junctions on a flexible platform, a network can be made to serve as a resistive strain sensor. Several experimental designs have been proposed using networks made of silver nanowires, carbon nanotubes and metal meshes with high sensitivities. However, there is limited theoretical understanding; the reported studies have taken the numerical approach and only consider rearrangement of nanowires with strain, while the critical break-make property of the sensor observed experimentally has largely been ignored. Herein, we propose a generic geometrical based model and study distortion, including the break-make aspect, and change in electrical percolation of the network on applying strain. The result shows that when a given strain is applied, wire segments below a critical angle with respect to the applied strain direction end up breaking, leading to increased resistance of the network. The percolation shows interesting attributes; the calculated resistance increases linearly in the beginning and at a higher rate for higher strains, consistent with the experimental findings. In a real scenario, the strain direction need not necessarily be in the direction of measurement, and therefore, strain value and its direction both are incorporated into the treatment. The study reveals interesting anisotropic conduction features; strain sensitivity is higher parallel to the strain, while strain range is wider for perpendicular measurement. The percolation is also investigated on direct microscopic images of metal networks to obtain resistance-strain characteristics and identification of current percolation pathways.  The findings will be important for electrical percolation in general, particularly in predicting characteristics and improvising metal network-based strain sensors.

\end{abstract}

\pacs{bhjh}

\maketitle

\section{Introduction}

Electrical transport in a metal wire network has been an area of keen interest for their fascinating percolation properties~\cite{balberg1984percolation, vzevzelj2012percolating, lebovka2018anisotropy} and extensive range of promising transparent conducting electrode applications.~\cite{ye2014metal,kumar2016evaluating,mutiso2013integrating} A strain sensor is an emerging application of metal wire networks with broader strain working range, extraordinary response and high transmittance, desired for robotics and health monitoring systems applications.~\cite{yao2014wearable, yamada2011stretchable,trung2016flexible} The metal network coupled with flexible substrate forms resistive-type strain sensor, its unknown value of applied strain can be determined, using the measured value of resistance. Various efforts have been devoted in the direction of fabricating metal network based strain sensors using silver nanowires, carbon nanotubes and template based metal meshes ~\cite{xu2012highly, amjadi2014highly} and their characteristics have been studied in detail. The studies exhibit that the resistance increases linearly for a lower strain, while non-linearly for higher strain values, resistance variation  in the parallel and perpendicular direction of strain are different~\cite{kim2015highly, gupta2018cosmetically} and sensitivity is higher by using binary width distribution.~\cite{duan2018highly} It is yet a challenge to improve their sensitivity and strain range significantly, for diverse applications. A comprehensive theoretical model is a need of the hour to explain their mechanism and discuss the pathway for their  further improvisation.

The electrical properties of metal networks are primarily investigated using the numerical method and have successfully calculated the sheet resistance with various parameters such as nanowire density,~\cite{vzevzelj2012percolating} length distribution~\cite{mutiso2013integrating}, anisotropy~\cite{lebovka2018anisotropy} etc. Besides, effective medium theory~\cite{o2016effective, he2018conductivity}, block matrix approach~\cite{kim2018systematic}, excluded volume percolation model~\cite{mutiso2012simulations}  have also been discussed for deeper insights. In this direction, we attempted to model it based on geometrical consideration and successfully obtained a relationship of sheet resistance and current carrying region with geometrical parameters of the network,~\cite{kumar2016evaluating,kumar2017current} which is found to be in experimental~\cite{darmakkolla2019morphology} and numerical agreement.~\cite{kim2018systematic} However, modeling of metal network based strain sensor is intensely involved, as it comprises distortion of the network with a given strain and then analyzing electrical transport of the distorted networks with the strain. Shengbo et al.~\cite{shengbo2018highly} have presented schematic representation for strain effect; however, theoretical or numerical calculations were not performed in the study. Amjadi et al.~\cite{amjadi2014highly} and Kim et al.~\cite{kim2015highly} developed 3-dimensional models of silver nanowire network and examined its resistance variation numerically during elongation. Yao et al.~\cite{yao2018characterizing} have very recently determined percolative coefficients, which are quantifiers of electrical transport properties,  using the strain sensor characteristics, exhibiting the interesting utility of strain studies for electrical percolation understanding. All these existing studies account for elongation and rearrangement of nanowires, however, do not include  breakdown of wire segments into the treatment. On the other hand, recent microscopic study by Gupta et al.~\cite{gupta2018cosmetically} reveals that wire segments, particularly for metal mesh, break during the strain and being half embedded in the elastic substrate, the broken wire segments reconnect at the same places on restoring strain, leading to the recovery of conductance for several such cycles. The present work pertains to the modeling of metal-network based strain sensor as a 2-dimensional graph; the application of strain distorts the network and breaks a few wire segments which in turn, modifies its electrical percolation significantly. Moreover, we calculate resistance and gauge factor as a function of strain and compare it for different strain directions and discuss some approaches for their improvement.
\begin{figure}[h]
\centering
\includegraphics[width=0.5 \textwidth ]{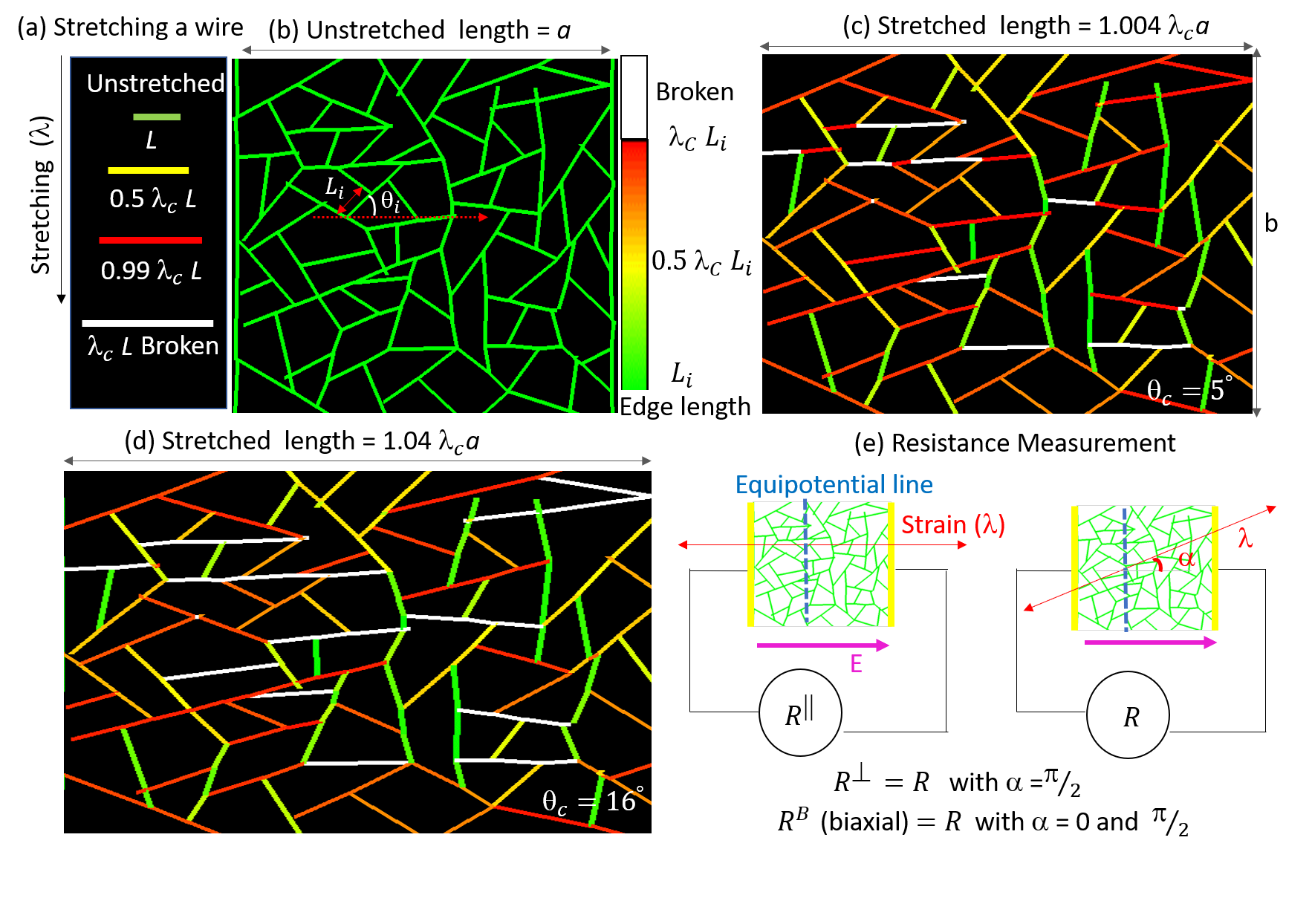}
\caption{(a) A single wire with length $L$ being elongated to $\lambda L$, the wire can elongate till a critical length $\lambda_C L$, beyond which, it ends up breaking. (b) Schematic representation of a metal network before strain. The strain is applied along X-axis (dotted red arrow), $L_i$ represents an edge length, and $\theta_i$ denotes its angle with respect to the strain direction. (c) The metal network is elongated to $1.004 \lambda_c a$, all the wire segments of network elongate, color represents their elongated lengths and few of them, marked in the white, break down. (d) Metal network elongated to $1.04 \lambda_c a$, wire segments elongate further and several more wire segments end up breaking. (e) Resistance measurement along various strain directions. $R^{\|}$ (left) is the resistance parallel to the direction of strain, $R$ (right) is the resistance measurement at an angle $\alpha$ with respect to the strain. $R^{\bot}$ is the resistance perpendicular to the strain and $R^{B}$ is the resistance measurement for a biaxial strain.}
\label{strain_sensor1}
\end{figure}
\section{Results and discussion}
To describe the mechanism of such sensors,  consider a random conducting network as a 2D graph~\cite{kumar2016evaluating} of size $a \times b$, where wire segments signify the edges and junctions denote nodes of the graph. We assume that the network elongates along the direction of strain and increase in resistance is primarily due to the breakdown of specific edges (wire segments).
A wire segment in a network breaks, if the experienced strain exceeds the critical strain for breakdown ($\lambda$\textgreater$\lambda_c$), here, critical strain, $\lambda_c$ depends on the strength of the wire (Fig. ~\ref{strain_sensor1}a). Further, we understand the role of strain in a network (Fig. ~\ref{strain_sensor1}b) with $\lambda$ being elongation along the horizontal direction. Fig. ~\ref{strain_sensor1}c-d represents network with length $\lambda $ = 1.004 and 1.04 times $\lambda_c$. Since all the edges have different angles w.r.t direction of strain and hence experience unequal strains. The strain experienced ($\lambda_E$) by an edge having angle $\theta_i$ w.r.t direction of strain is $\lambda_{E}= \lambda cos\theta_i $. The wire segments break if $\lambda_E$\textgreater$\lambda_c$, in other words, the wire segments with an angle lesser than critical angle end up breaking ($\theta_i$ \textless $\theta_C$). The critical angle can be defined as

\begin{equation}
\theta_C = \cos^{-1} \frac{\lambda_c}{\lambda}
\label{thetab}
\end{equation}
As an example, with the strain values 1.004 $\lambda_c$ and 1.04 $\lambda_c$, edges with angle less than 5$^0$ and 16$^0$ respectively end up breaking. To model its electrical resistance , lets begin with, $\alpha =0$ i.e. resistance is measured along the direction of the strain as $R^{\|}$ (see Fig.~\ref{strain_sensor1}e left). As, edge length $<<$ network size, one can assume potential drops uniformly between the verticle electrodes, and equipotential lines are perpendicular to the electric field. Let  $L_i$ be the length of wire segment (edge) placed in the electric field $E$ at an angle $\theta_i$. The potential difference across the edge, $V_i$, therefore depends on the orientation of edge as

\begin{equation}
\label{Vr}
V_{i} = E L_{i} cos \theta_{i}
\end{equation}

As wire segments with an angle lesser than  $\theta_C$ are broken with the starin, thus the average potential of all unbroken wire segments can be calculated as
\begin{equation}
\label{Vr_avg}
V_{am} = \frac{ \int_{\theta_C}^{\frac{\pi}{2}} EL_i cos\theta_i d\theta_i} {\int_{\theta_C}^{\frac{\pi}{2}} d\theta_i} = EL \frac{1 - sin\theta_C}{\frac{\pi}{2}-\theta_C}
\end{equation}

\begin{figure}[h]
\centering
\includegraphics[width=0.5\textwidth ]{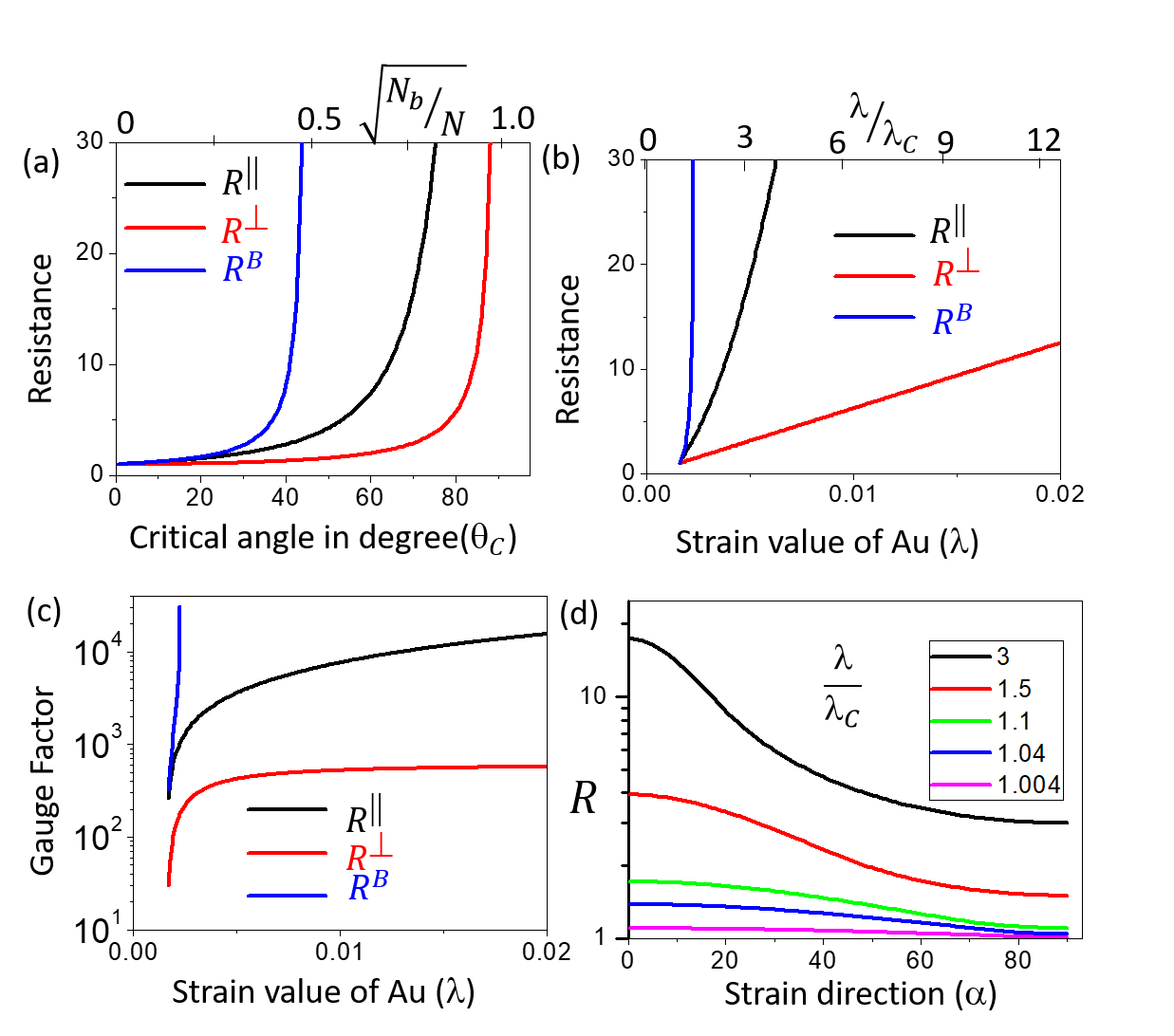}
\caption{ (a) Comparision of variation in resistance parallel to strain, perpendicular to strain and biaxial strain with respect to the density of broken wire segments in upper axis and the critical angle of the breakdown in the upper axis. (b) Variation of resistance with respect to strain values with scale $\frac{\lambda_c}{\lambda}$ in upper axis and typical values for Au network in the lower axis. (c) Variation of gauge factor for Au network for the increasing strain. (d) Variation in resistance, $R$ with respect to strain angle at different strain values}
\label{strain_sensor3}
\end{figure}

Say, $\rho$ is the resistivity, $w$ is the width, and $t$ is the thickness of the wire segment (edge), then the mean resistance of an individual edge is $ \frac{\rho L_{am}}{w t}$. Hence, mean current, $I_{am}$ passing through an individual edge can be written as

\begin{equation}
\label{ir2}
I_{am}= \frac{E wt} {\rho} \frac{1-sin\theta_C}{\frac{\pi}{2}-\theta_C}
\end{equation}

Consider, $N$, $N_b$ and $N_u$ as the edge density of total, broken and unbroken wire segments respectively. The current passing across an equipotential line, $I_{eq}$, depends on the current carried by a single edge and a total number of unbroken edges on the equipotential line which equals $\sqrt{N_u} \, b $ by symmetry arguments.

\begin{equation}
\label{i}
I_{eq}= \frac{E wt} {\rho} \frac{1-sin\theta_C}{\frac{\pi}{2}-\theta_C} \sqrt{N_u}b
\end{equation}

Using, $E = \frac{V}{a} $ and applying Ohm's law, resistance, $R^{\|}$ can be written as
\begin{equation}
\label{R11}
R^{\|} = \frac { \rho} {wt\sqrt{N_u}} \frac{\frac{\pi}{2}-\theta_C} {1-sin\theta_C} \frac{a}{b}
\end{equation}
As the wire segemnts with angles  up-to angle $\theta_C$ are broken, hence $ \sqrt{N_u} = \frac{\frac{\pi}{2} -\theta_C}{ \pi /2} \sqrt{N} $. If $R^0$ is the resistance, in the unstreched state.
\begin{equation}
\label{R311}
R^{\|} = \frac{R^{0}} {1-sin\theta_C}
\end{equation}
Using $\theta_C$ = $ \cos^{-1} \frac{\lambda_c}{\lambda}$ from Eq. ~\ref{thetab}, above can written as

\begin{equation}
\label{R51}
R^{\|} = \frac{R^{0}} {1-sin(cos^{-1} \frac{\lambda_c}{\lambda})} =\frac{R^{0}} {1- \sqrt{ 1- \big(\frac{\lambda_c}{\lambda}}\big)^2}
\end{equation}
Using the first term of series expension, for lower strain values, the resistance shows linear increase as
\begin{equation}
\label{R5178}
R^{\|} \approx 2 R^{0} \frac{\lambda}{\lambda_c}
\end{equation}

Now, if one is measuring resistance perpendicular to the direction of strain, the average potential difference can be written as $V_{am}^{\bot} = \frac{ \int_{0}^{\frac{\pi}{2}-\theta_C} EL_i cos\theta_i d\theta_i} {\int_{0}^{\frac{\pi}{2}-\theta_C} d\theta_i}$ and the resistance,$R^{\bot}$ can be calculated as
\begin{equation}
\label{R311p}
R^{\bot} = \frac{R^{0}} {cos\theta_C} =
{R^{0}} \frac{\lambda}{\lambda_c}
\end{equation}

Further, consider a generic case having an angle of strain, $\alpha$ with respect to the resitance measurement direction (Fig. ~\ref{strain_sensor1}e right). As derived in supporting information, $R$ depends on value of $\alpha +\theta_C$. For $\alpha +\theta_C <\frac{\pi}{2}$, based on Eq. S8,
\begin{equation}
\label{Rx1}
R =
\frac{R^{0}} { 1-cos\alpha \, sin\theta_C}
\end{equation}
and for $\alpha +\theta_C > \frac{\pi}{2}$, based on Eq. S13,
\begin{equation}
R =
\frac{R^{0}} { sin\alpha \, cos\theta_C}
\label{Rx11}
\end{equation}
In this way, one can calculate resistance for any given value of strain and its direction, which is necessary in most of the biological and robotics applications, where the strain direction need not necessarily be in the direction of measurement. Similarly, any other experimentally interesting case, such as biaxial strain~\cite{ryu2015extremely} can also be studied by the treatment. For biaxial strain, with $\lambda$ being starin along both the direction, the resistance can be expressed as
\begin{equation}
\label{Rb}
R^{B} = \frac{R^{0}} {cos\theta_C - sin\theta_C} =
\frac{R^{0}} { \frac{\lambda_c}{\lambda} - \sqrt{ 1- \big(\frac{\lambda_c}{\lambda}}\big)^2}
\end{equation}

\begin{figure*}
\centering
\includegraphics[width=1.0 \textwidth ]{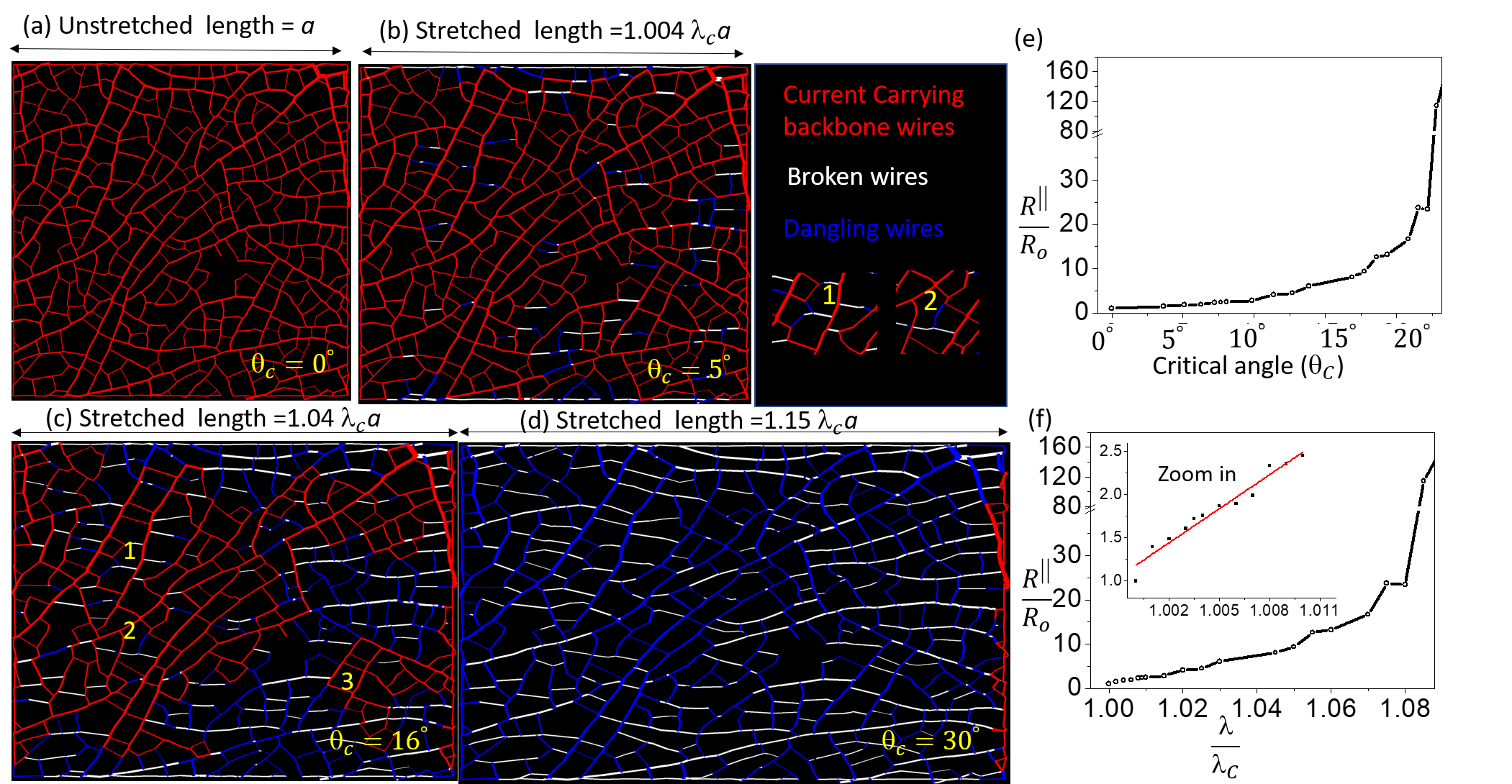}
\caption{ (a-d) Classification of current carrying backbone, broken and dangling regions of a metal network for different strain values using image analysis. Regions 1 and 2 are zoomed in for clarity, region 3 is an artifact of backbone as it makes circular loop. (e-f) Variation of numerically calculated resistance with critical angle and increasing strain, inset illustrates linearity at lower strain values. }
\label{image_strain}
\end{figure*}

Based on the above calculations, Fig. ~\ref{strain_sensor3}a represents variation in resistance with an increase in critical angle in lower axis and edge density of broken wires, in the upper axis as $\frac{\theta_C}{\pi/2} = \sqrt \frac{N_b}{N}$. The increase in resistance is gradual in the beginning and rises abruptly for larger critical values or broken edge densities. Fig. ~\ref{strain_sensor3}b shows the results in terms of critical strain ($\lambda_c$), and for typical strain value for Au. Here, $\lambda_c$ for Au considered is approximately $1.6 \times 10^{-3}$ calculated using $\lambda_c= \frac{G}{Y}$ by substituting typical Yield strength $(G)$ of 50 to 200 $MPa$ and Young Modulus $(Y)$ 8 GP for Au.~\cite{wu2005mechanical} It is clear, for resistance measurement along the direction of strain (black curve), the resistance increases linearly for lower strain, while it increases drastically for higher strain values. Further, the working range of the strain sensor in all cases is different. For parallel strain (black curve), resistance shoots up at relatively lower strains as compared to the perpendicular strain (red curve); thus higher strain measurement may be more appropriate to measure in the perpendicular direction. For biaxial strain, resistance shoots up for relatively smaller strains and thus such sensors may not be suitable to measure higher biaxial strains. The phenomena can be explained by considering the importance of the current carried by the broken wire segments. For perpendicular strain, broken wire segments are nearly on equipotential lines carrying negligible current; thus, there is very less effect on resistance after their breakdown. On the other hand, for parallel strain, broken wire segments are nearly in the direction of the electric field carrying maximum current; thus their breakdown leads to more increase in overall resistance. For biaxial strain, both kinds of wire segments are broken, leading to their maximum increase in resistance. Gauge factor defined as $\frac{\bigtriangleup R}{\lambda}$ can also be derived easily and shown in Fig.~\ref{strain_sensor3}c. In this way, one can choose and understand the limitations of measurement mode based on sensitivity and working strain range. The decrease in resistance with increasing strain angle from parallel to perpendicular is also clear in (Fig. ~\ref{strain_sensor3}d) and can be used to measure resistance change for arbitrary strain direction.

Further, we demonstrate the pertinence of the model on a microscopic image of a metal wire network as an example (Fig. ~\ref{image_strain}). The analysis involves converting a metal network image into 2- dimensional graph using image analysis, distorting the graph for various strain values and numerically calculating their percolating path and corresponding sheet resistance values. Only the current carrying backbone is shown in the initial image for clarity (Fig. ~\ref{image_strain}a);  with the applied strain, edges with projection angle lower than the critical angle, break down (white), and some edges become dangling with no current (blue), and current flows in the rest of current carrying backbone region (red). Interestingly, the current carrying backbone (red) gradually decreases for lower strain values, $\lambda = 1.004\lambda_c$ (Fig. ~\ref{image_strain}b) while for higher strain values, $\lambda = 1.04\lambda_c$ (Fig. ~\ref{image_strain}c), the current carrying backbone forms bottleneck (Fig. ~\ref{image_strain}d), in the end, at $\lambda = 1.15\lambda_c$, there is no current carrying path. The resistance is determined for all the networks,  using the two-point resistance method, based on solving Kirchhoff's law in the network.\cite{ kumar2017current} Resistance is found to increase linearly for lower strain (see inset of Fig. ~\ref{image_strain}f) and increases with the higher rate for higher strain values before the breakdown of percolation pathway (see Fig. ~\ref{image_strain}f). In this way, interesting percolation scenareos can be seen to work in different strain regimes. 

The model is generic and can be applied to many special cases of metal networks and different strain modes. If the deposited wire networks have good adhesion with the substrate, the nanowires cannot individually move, and network as a whole elongates, as discussed in the model. The technique can compare various microscopic images of metal networks, and help in the selection of the optimum one, for strain sensor applications. Additionally, the model can be utilized in understanding the robustness of electrical percolation in the networks for stretchable electronics applications~\cite{guo2014highly}, where a minimum variation of resistance with a strain  is desired. The study brings out important features of anisotropic conduction and its implications in strain sensor characteristics. The present day strain sensor devices only measure strain values, the model may be an important step towards realizing strain sensor devices, which could measure strain value and its direction simultaneously. Note that the analytical treatment is based on applying effective medium theory, which may not hold good at very sparse network in case of very high strains; image analysis based numerical approach discussed here, should be applied on bigger images in such  contexts.  For a more accurate and detailed picture, future work should improvise the model by calculating dangling regions with strain, introducing corrections for a sparse network, and introducing self-healing effects~\cite{kumar2019self} and quantum conductance of interconnects. The present model estabilishes important modeling ideas, which  will be helpful in approaching much complex scenareos. 
\section{Conclusion}
In conclusion, a geometrical model is presented to understand the impact of strain on the electrical percolation properties of metal wire networks. The analysis explains that  strain applied to the network leads to a preferential breakdown of certain wire segments based on their angle with respect to the strain direction, which increases their resistance. The calculated resistance increases linearly for the lower strain values and abruptly for higher strains,  consistent with experiments.  The electrical percolation manifests interesting anisotropic effects; the variation in resistance with strain is higher along the direction of strain, while the strain range is higher if the measurement is performed perpendicular to the strain. The study devised a numerical approach to obtain the percolating pathways and strain sensor characteristics from microscopic images of metal networks. The analysis suggests that the current carrying region decreases gradually for lower strains, while forms a bottleneck pathway at higher strains and finally collapse completely at very high strain values. The proposed model and its interesting findings will be useful in addressing various electrical percolation problems and improvising strain sensor devices.  

See supplementary material for detailed calculations of resistance variation for strain in an arbitrary direction.

\section*{Acknowledgements}
The author acknowledges his Ph.D. mentor Professor Giridhar U. Kulkarni for valuable suggestions, encouragement, and proposing the problem based on lab experimental observations and acknowledges the Department of Science and Technology, India for financial support.


\bibliography{references}

\end{document}